\documentclass[prl,twocolumn,showpacs]{revtex4}
\usepackage[usenames]{color}
\usepackage{epsf}
\usepackage{epsfig}
\usepackage{graphicx}
\usepackage{latexsym}
\usepackage{bm}
\begin{document}

\title{
Localization of Inner-Shell Photoelectron Emission and Interatomic
Coulombic Decay in Ne$_2$ }
\author{
K.~Kreidi$^{1,2}$, T.~Jahnke$^1$, Th.~Weber$^{3}$,
T.~Havermeier$^1$, R.~E.~Grisenti$^{1,4}$, X.~Liu$^5$,
Y.~Morisita$^6$, S.~Sch\"ossler$^1$, L.~Ph.~H.~Schmidt$^1$,
M.~Sch\"offler$^1$, M.~Odenweller$^1$, N.~Neumann$^1$,
L.~Foucar$^1$, J.~Titze$^1$, B.~Ulrich$^1$, F.~Sturm$^1$,
C.~Stuck$^1$, R.~Wallauer$^1$, S.~Voss$^1$, I.~Lauter$^1$,
H.~K.~Kim$^1$, M.~Rudloff$^1$, H.~Fukuzawa$^5$, G.~Pr\"umper$^5$,
N.~Saito$^6$, K.~Ueda$^5$, A.~Czasch$^1$,
O.~Jagutzki$^1$, H.~Schmidt-B\"ocking$^1$, S.~K.~Semenov$^7$, N.~A.~Cherepkov$^7$\\
and R. D\"orner$^{1}$}

\email{doerner@atom.uni-frankfurt.de}
\address{
$^1$ Institut f\"ur Kernphysik, J.~W.~Goethe Universit\"at, Max-von-Laue-Str. 1, 60438 Frankfurt, Germany \\
$^2$ DESY, Notkestrasse 85, 22607 Hamburg, Germany\\
$^3$ Lawrence Berkeley National Laboratory, Berkeley CA 94720, USA\\
$^4$ Gesellschaft f\"ur Schwerionenforschung, Planckstr. 1, 64291 Darmstadt, Germany \\
$^5$ Institute of Multidisciplinary Research for Advanced Materials,
Tohoku University, Sendai 980-8577, Japan\\
$^6$ National Metrology Institute of Japan, AIST, Tsukuba 305-8568,
Japan\\
$^7$ State University of Aerospace Instrumentation, 190000 St.
Petersburg, Russia }

\begin{abstract}
We used Cold Target Recoil Ion Momentum Spectroscopy (COLTRIMS) to
investigate the decay of Ne$_2$ after K-shell photoionization. The
breakup into Ne$^{1+}$~/~Ne$^{2+}$ shows interatomic Coulombic decay
(ICD) occurring after a preceding atomic Auger decay. The molecular
frame angular distributions of the photoelectron and the ICD
electron show distinct, asymmetric features, which imply
localization of the K-vacancy created at one of the two atomic sites
of the Ne$_2$ and an emission of the ICD electron from a localized
site. The experimental results are supported by calculations in
frozen core Hartree-Fock approach.
\end{abstract}
\maketitle

Are inner-shell holes in homonuclear diatomic molecules localized at
one of the atoms or delocalized over the two equivalent sites? This
question has been discussed highly controversial in literature for
more than 35 years now (see e.g.
\cite{Snyder71jcp,Bagus72jcp,kintop91pra,Thiel03,Ehara06,dill78prl,Glans96,Semenov06jpb,Bjornholm00prl,chen89pra,Rolles05}
and our discussion below). Here we report on an experiment answering
that question for the Ne$_2$ Van-der-Waals-molecule. Intuitively
strong arguments for both opinions may be found: the K-shell wave
function is very tightly confined to the nuclei and the overlap
between inner-shell orbitals at different atoms of a molecule is
usually negligibly small \cite{Kosugi03cp}. Hence, the geometry of
the problem suggests to think of individual localized atomic wave
functions. The symmetry of the problem however suggests the
opposite: both sites of the diatomic molecule are indistinguishable
and therefore the total molecular wave function has to have well
defined $gerade$ ($g$) or $ungerade$ ($u$) symmetry. In order to
construct the molecular many-body wave function it seems natural to
employ only symmetry adapted single electron orbitals. A core level
hole would then have well defined $g$ or $u$ symmetry and hence be
delocalized over the two sites. This approach is used in today's
state of the art theory concerning the photoionization and decay of
Ne$_2$ \cite{Scheidt}.

To address this question of a possible localization in a quantum
mechanically meaningful and experimentally accessible way one has to
relate it to a measurable observable. Prime candidate is the energy
of the state measured by photoelectron spectroscopy. Theoretical
claims for core hole localization based on the energy date back to
classic works of Snyder \cite{Snyder71jcp} and Bagus
\cite{Bagus72jcp}. They showed that allowing symmetry broken - i.e.
localized - basis states for inner-shell vacancies in a Hartree-Fock
calculation lowers the energy and yields better agreement with the
experiment. It was later argued that this is a peculiarity of the
Hartree-Fock approach \cite{kintop91pra}. More sophisticated
calculations on N$_2$ today very well reproduce the experimentally
observed energy splitting of about 100~meV between the $1\sigma_g$
and $1\sigma_u$ core-ionized states \cite{Thiel03,Ehara06}. This is
generally taken as evidence for the delocalized character of the
inner-shell hole.

As an alternative observable being sensitive to core hole
localization the angular distribution of the photoelectrons in the
laboratory system was suggested \cite{dill78prl}. Corresponding
experiments on nitrogen molecules show good agreement with
calculations using delocalized orbitals \cite{Semenov06jpb}. This
conclusion is also supported by resonant soft X-ray emission
experiments on O$_2$ \cite{Glans96}. There the parity selection
rules indicate the symmetry and with it the delocalization of the
core-excited state. In contrast to that partial localization is
shown in experiments on singly substituted $^{14,15}$N$_2$
\cite{Rolles05}. Furthermore, experiments exciting an inner-shell
electron inducing fast dissociation and observing the Auger electron
emitted during this dissociation show a clear Doppler-shift in the
Auger peak. This proves that they are emitted from a localized
source flying towards to or away from the observer
\cite{Bjornholm00prl,Golovin97prl}.

\begin{figure}
  \begin{center}
  \vspace*{+3mm}
  \includegraphics[width=8cm]{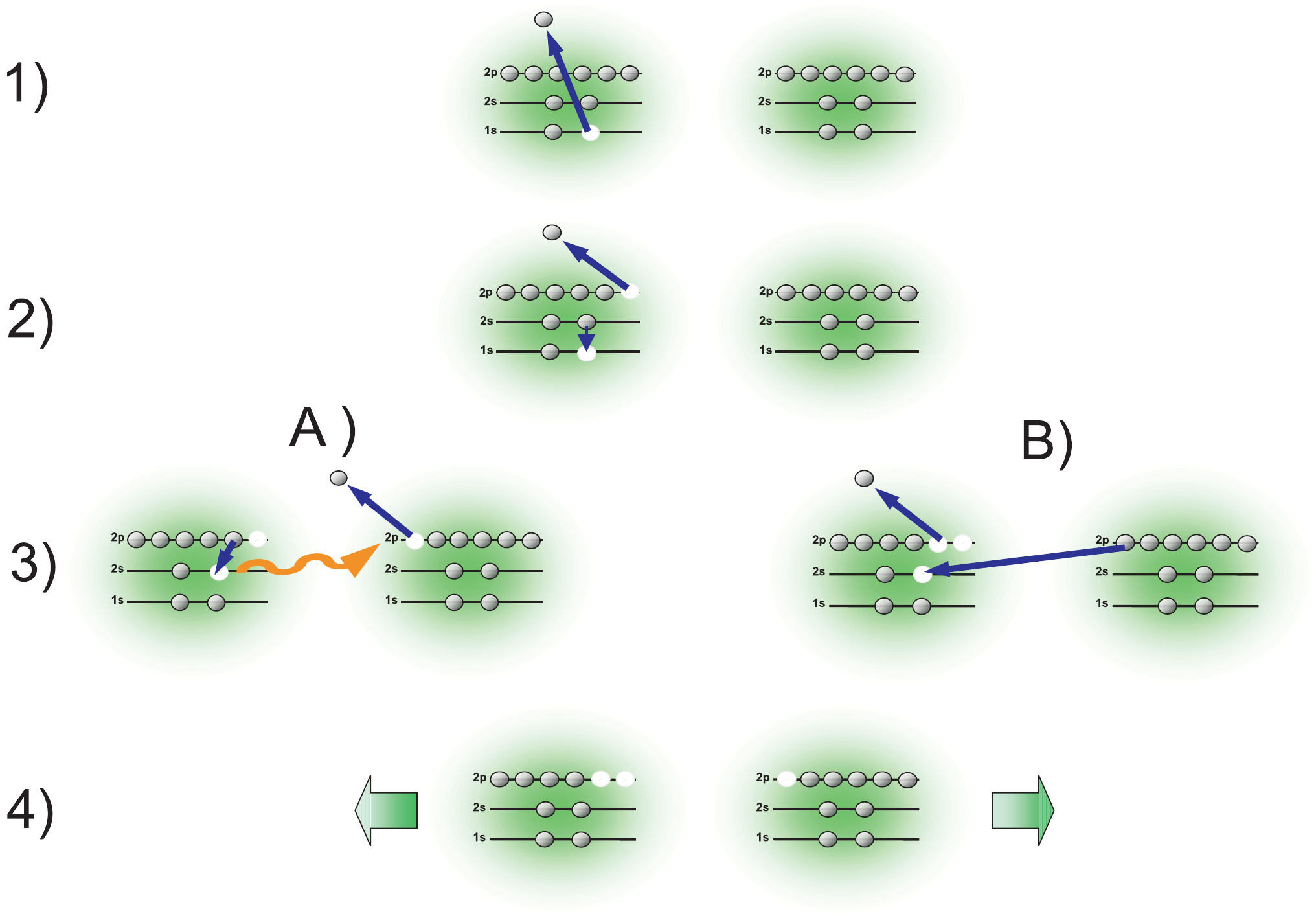}\\
  \caption{$1s$ photoionization (1) leads to an Auger decay resulting in a dicationic one-site state of Ne$_2$ (2).
  The Auger decay is followed by interatomic Coulombic decay (3)
  where the ion's excitation energy is emitted via a virtual photon transfer (A) or
  via electron transfer (B). Ending up in a state where both atoms
  are charged positively, the dimer fragments back-to-back in a Coulomb explosion
  (4).}
  \end{center}
  \label{Figure1}
\end{figure}

In the present work we use an even more sophisticated probe for
localization which is the electron angular distribution in the
\emph{body-fixed frame} of the molecule
\cite{Saito05jpb,Adachi07,Golovin97prl}: we investigate the angular
distribution of a photoelectron and an electron emitted via
interatomic Coulombic decay (ICD) \cite{Cederbaum97} following Auger
decay \cite{Santra03} for fixed molecular orientations. If the
electron is emitted from a delocalized source its angular
distribution will be symmetric with respect to the two atoms of the
molecule (see e.g. \cite{Pavlychev98prl,Weber01jpb}). If, however,
it is emitted from one site, then the angular distribution can show
a strong asymmetry due to interference of the electron-wave that is
multiply scattered in the molecular potential. For example, a strong
forward focussing by the atomic neighbor can occur \cite{Poon84}. In
a homonuclear diatomic molecule such an asymmetry however can only
be observed for an asymmetric breakup of the molecule into two
fragments of different charge which make the two ends of the
molecule experimentally distinguishable. A first experiment along
this line on N$_2$ found no asymmetry with respect to the
N$^{1+}$/N$^{2+}$ fragments \cite{Weber01jpb}. Here we investigate
K-shell ionization of Ne$_2$ instead of N$_2$ for several reasons.
Firstly, K-shell ionized Ne$_2$ fragments into the asymmetric
breakup channel Ne$^{1+}$~/~Ne$^{2+}$. Secondly the K-shell radius
of Ne$_2$ is 30 times smaller than the internuclear distance $R$ of
5.86~a.u. The bond is purely Van-der-Waals, the binding energy is
only 3~meV and the $g/u$ splitting is negligibly small compared to
the natural line width. All this might make it plausible to think of
Ne$_2$ as two Neon atoms sitting close together with their K-shells
being independent of each other. Thirdly however the opposing
symmetry argument in favor of a delocalized description of all
electrons by symmetry adapted wave functions of well defined $g$ or
$u$ parity is as valid for Ne$_2$ as it is for N$_2$. Hence the key
arguments keeping the question of possible localization open today
in covalently bound molecules hold for Van-der-Waals dimers, as
well.

Santra \emph{et al.} \cite{Santra03} theoretically investigated the
decay of Ne$_2$ after $1s$ ionization. They suggested that one-site
Auger final states Ne$^{2+}$(2s$^{-1}$2p$^{-1}$)~/~Ne decay further
to a two-site state Ne$^{2+}$(2p$^{-2}$)~/~Ne$^{1+}$(2p$^{-1}$) via
ICD. Fig. 1 shows a sketch of the different pathways of decay
leading to this set of reaction products. Both decay routes start
with K-shell ionization. In a localized scenario it is followed by
an Auger decay of the same atom of the dimer yielding a vacancy in
an inner valence shell (Ne$^{2+}$(2s$^{-1}$2p$^{-1}$)~/~Ne). Now two
different channels of ICD may occur depending on the parity of the
doubly charged states \cite{Santra01,Jahnke07}. Pathway A) shows ICD
via virtual photon exchange, B) depicts the competing process
involving the transfer of an electron. As both A) and B) lead to a
triply charged species, Ne$_2$ fragments in a Coulomb explosion as a
final step. The existence of ICD with its two center nature is the
key feature in Ne$_2$, making it possible to trace core hole
localization through all steps of the decay of Ne$_2$. The sketch of
the decay in Fig.~1 shows fully localized electrons and holes. Thus
it may seem that the pure existence of ICD already proofs a
localization. This is not true however. While most discussions on
ICD on an intuitive level use figures similar to our Fig. 1, all
actual calculations on ICD, including the ones performed by Santra
for the present system \cite{Santra03}, work with molecular wave
functions of well defined $g$ or $u$ symmetry, i.e. they assume
complete delocalization. The ICD calculations using symmetry adapted
wave functions yield excellent agreement of the experimentally
observed energies of the states and of the energy of the ICD
electrons \cite{Jahnke04prl,Scheidt}. These calculations report ICD
electron energies separately for $g$ and $u$ symmetry again
highlighting the delocalized character of the calculations
\cite{Scheidt}.

%The same calculations, however, could be used for the case of
%localization by coherently adding the amplitudes for g and u
%transitions.
\begin{figure}
  \begin{center}
  \vspace*{+3mm}
  \includegraphics[width=7cm]{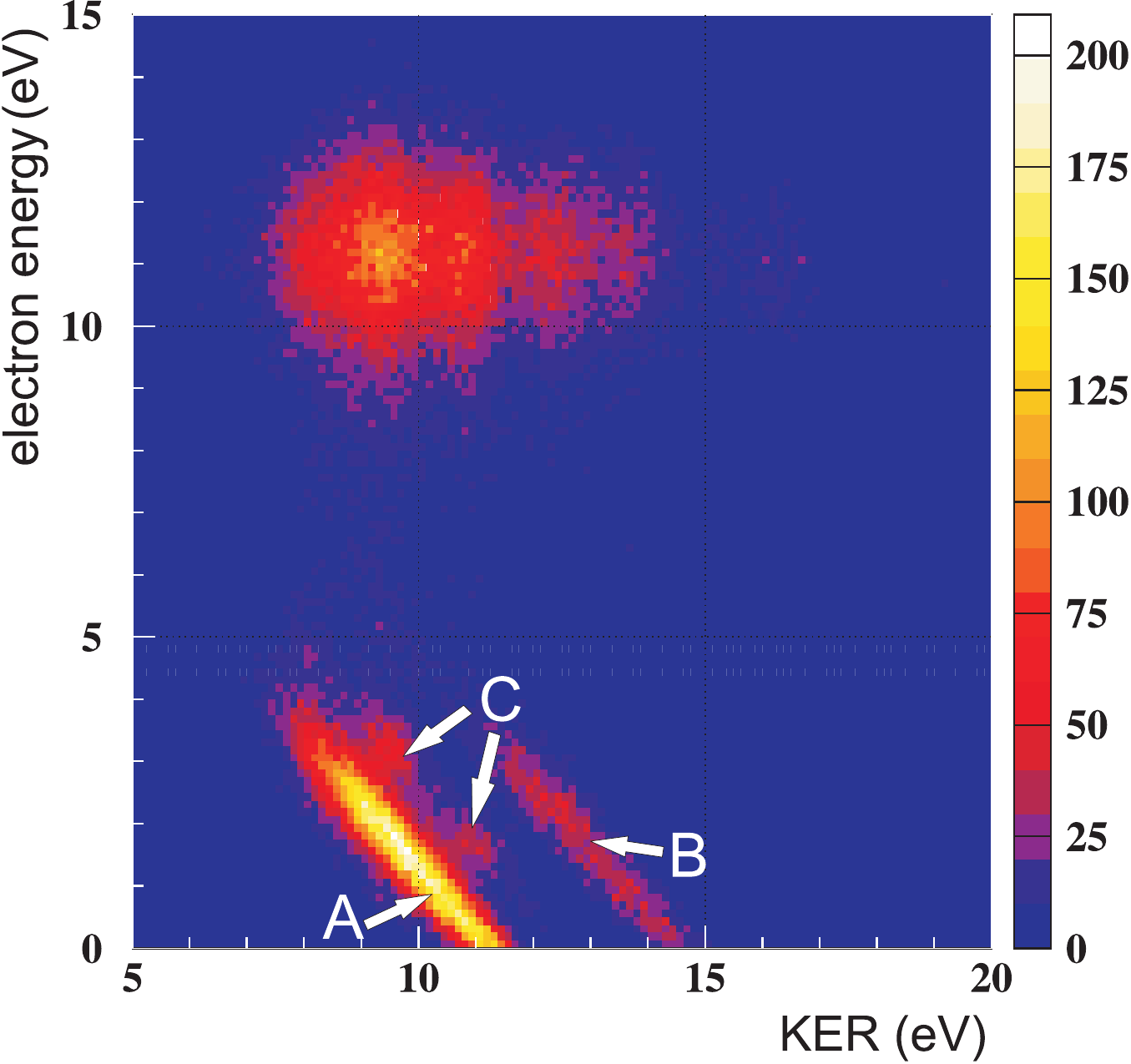}\\
  \caption{The kinetic energy of the ionic fragments in dependence of the electron energy. The photon
  energy is $h\nu=881.2$~eV resulting in an energy of the $1s$ photoelectron of 11~eV.
  The two diagonal lines are a clear evidence for interatomic
  Coulombic decay (ICD). Line A) is produced via the virtual photon exchange
  yielding a sum energy of 11.1~eV. Line B) (with a sum energy of 14.3~eV) occurs as ICD via electron
  transfer happens. The origin of channel C), showing two separate islands, is not quite clear.}
  \end{center}
  \label{fig:Figure2}
\end{figure}

The experiment was performed at beamline UE56/1-SGM of the Berlin
Synchrotron (BESSY) using the COLTRIMS technique
\cite{ullrich03rpp}. As a source for Ne$_2$ a supersonic jet that
was precooled to 160~K was employed. It was crossed with the photon
beam. Products from the photoreaction were guided by an electric
field of $20$~V/cm and a magnetic field of $6$~Gauss towards two
channel plate detectors with delayline readout \cite{jagutzki02nim}.
Electrons up to 12~eV and ions up to $10$~eV were detectable with a
solid angle of $4\pi$.

Fig. 2 shows our experimental results for the measured electron
energies and the sum of the kinetic energies of the ionic fragments
(KER, kinetic energy release). With a photon energy of
$h\nu=881.2$~eV, events located at an electron energy of 11~eV
correspond to photoelectrons from Ne $1s$ ionization. Electrons
originating from ICD are identified as diagonal lines A) and B). As
the sum of the energy of the ionic fragments and the ICD electron is
a constant, these diagonal lines with a slope of $-45^{\circ}$ are a
clear evidence for ICD as shown in \cite{Jahnke04prl}.

The ICD channels, labeled A) and B) in Fig. 2, have a different KER
which shows that the decay occurs at different internuclear
distances. Similar to the findings in \cite{Jahnke07} both channels
are created in an IC decay with the symmetry of the involved states
and the difference in kinetic energy of the ions implying the
occurrence of the two different contributions to ICD: channel A)
with a sum energy of 11.1~eV represents the decay from the inner
valence excited one-site state
Ne$^{2+}$~(2s$^{-1}$2p$^{-1}$)~[$^1P$]~/~Ne~[$^1S$] to the two-site
state Ne$^{2+}$~(2p$^{-2}$)~[$^1D$]~/~Ne$^{1+}$~(2p$^{-1}$)~[$^2P$].
This so called 'direct' IC decay happens via an exchange of a
virtual photon as sketched as pathway A in Fig. 1. The measured KER
of $\sim$~8~eV to $\sim$~11~eV corresponds to an internuclear
distance of $\sim$~3.6~{\AA} to $\sim$~2.6~{\AA}
\cite{Weber04nature}. The mean value of this range is close to
3.05~{\AA} which is the distance of the neon atoms in the ground
state of Ne$_2$ and thus consistent with corresponding results in
\cite{Jahnke07}. Channel B) with a sum energy of KER + electron
energy~=~14.3~eV occurs at much higher KER ($\sim$~11~eV to
$\sim$~14~eV) which is equivalent to a internuclear distance of only
$\sim$~2.6~{\AA} to $\sim$~2.0~{\AA}. Here the one-site state
Ne$^{2+}$~(2s$^{-1}$2p$^{-1}$)~[$^1P$]~/~Ne~[$^1S$] decays into the
two-site state
Ne$^{2+}$~(2p$^{-2}$)~[$^3P$]~/~Ne$^{1+}$(2p$^{-1}$)~[$^2P$] after
electron transfer, as shown by pathway B in Fig. 1. The decay is
described by the 'exchange' part of the electron-electron Coulomb
matrix element. In this case the spatial overlap of the involved
wave functions is the crucial contribution to the decay probability.
Therefore pathway B is suppressed at large internuclear distances
where pathway A is still open \cite{Jahnke07}. Beside channels A and
B our experimental results also show two small islands, labeled as C
in Fig. 2. The sum energy of KER + electron energy~=~12.1~eV
indicates the decay from the dicationic state
Ne$^{2+}$~(2s$^{-2}$)~[$^1S$]~/~Ne~[$^1S$] to the two-site state
Ne$^{2+}$~(2s$^{-1}$)(2p$^{-1}$)~[$^3P$]~/~Ne$^{1+}$~(2p$^{-1}$)~[$^2P$].
The KER of this decay is $\sim$~11~eV which is equivalent to
$R$~=~2.6~{\AA}. This channel also represents a spin-flip ICD where
electron transfer may play a role but until now it is not clear why
it shows two separate islands instead of a complete diagonal line.

\begin{figure}
  \begin{center}
  \vspace*{+3mm}
  \includegraphics[width=7cm]{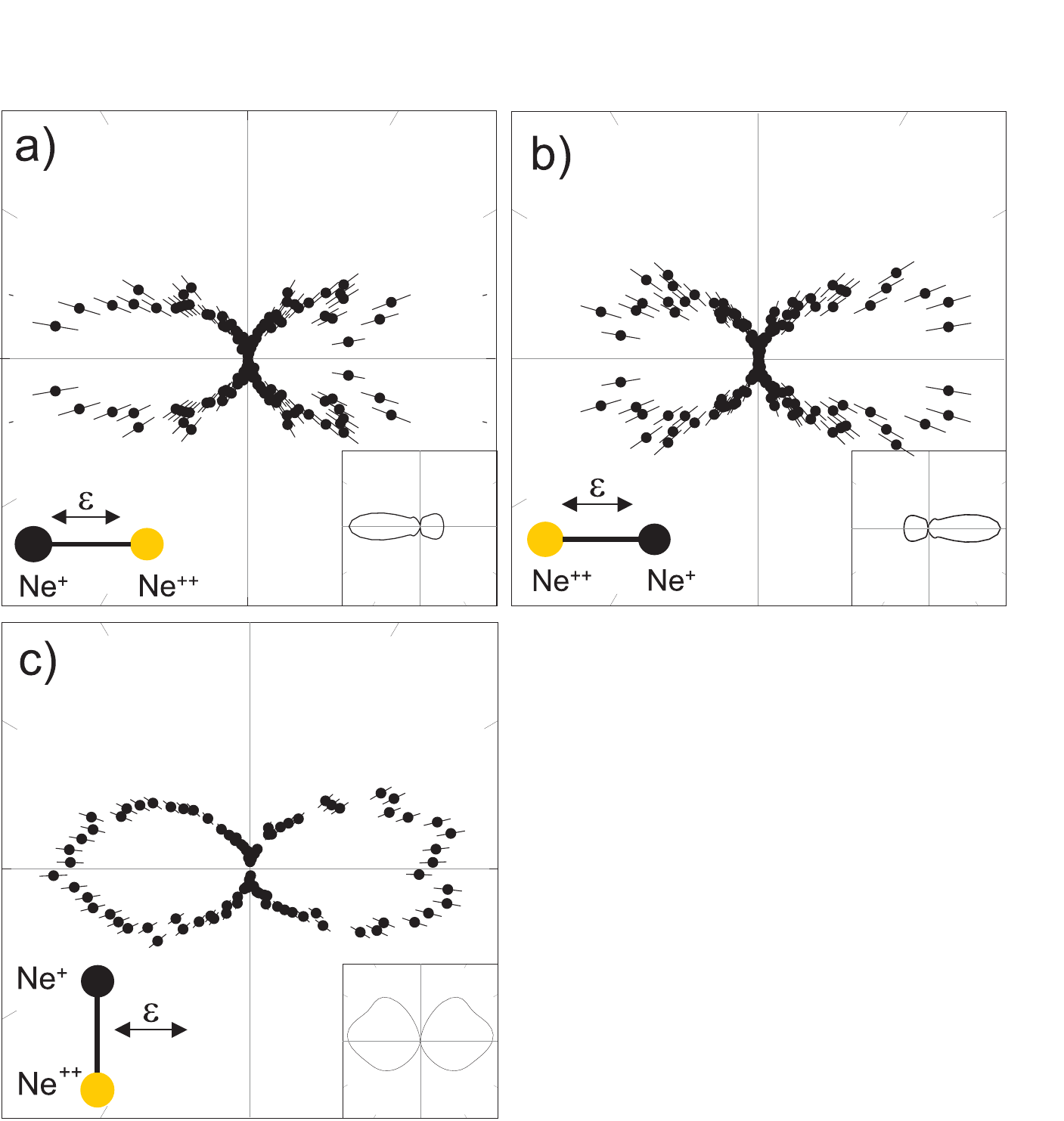}\\
  \caption{Angular distribution of the 11 eV $1s$ photoelectron in dependence of the
  orientation of the dimer axis and the direction of the polarization vector $\varepsilon$ (horizontal).
  In a) and b) the
  dimer is aligned parallelly to the polarization vector within an angle of
  $\pm~3^{\circ}$. In c) the plane is defined by the dimer axis and the polarization of the light. The dimer is aligned
  perpendicularly to the polarization vector and the photoelectron is fixed relative to the plane within an angle of
  $\pm~10^\circ$. The asymmetry, being a result of core hole localization, is clearly visible in the
  experimental data (circles). Solid line: frozen core Hartree-Fock calculation assuming a localized emission. }
  \end{center}
  \label{Figure3}
\end{figure}

Fig. 3 shows the angular distribution of the emitted photoelectron
in the body fixed frame of Ne$_2$. Clearly, the photoelectron
angular distribution is asymmetric - the photoelectron is preferably
emitted towards the singly charged dimer fragment. To validate our
findings Fig. 3(a) and 3(b) show the results where the doubly
charged ion is emitted in two opposite directions within the
laboratory frame. As a support for our interpretation we compared
the experimental angular distributions with a theoretical prediction
for the case of a completely localized electron. The calculation was
performed within the Hartree-Fock approximation using the method
described in \cite{Semenov00}: at first the ground state of Ne$_2$
was computed in order to obtain the initial $1\sigma_g$ and
$1\sigma_u$ wave functions. The photoelectron wave function was
calculated in the frozen core Hartree-Fock approximation.
Localization of the initial hole after photoionization on the right
or left atom ($|r\rangle$ or $|l\rangle$ states) was introduced by
taking a linear combination of the symmetry adapted $1\sigma_g$ and
$1\sigma_u$ wave functions, $|r\rangle = (|1\sigma_g\rangle +
|1\sigma_u\rangle$)$/\sqrt{2}$ and $|l\rangle = (|1\sigma_g\rangle -
|1\sigma_u\rangle$)$/\sqrt{2}$. The full line down right in Fig.
3(a), 3(b) and 3(c) shows the prediction for the assumption of fully
localized emission of the photoelectron from the doubly charged part
of Ne$_2$. It nicely resembles the asymmetric shape of the measured
distribution, overestimating the total magnitude of the asymmetry.
Fig. 3(c) shows the same distribution with the dimer being oriented
perpendicularly to the polarization axis of the photon beam. As
expected no left/right asymmetry is visible in that case, but a
small up/down asymmetry, which is again overestimated by our
calculations that imply complete localization.

\begin{figure}
  \begin{center}
  \vspace*{+3mm}
  \includegraphics[width=8cm]{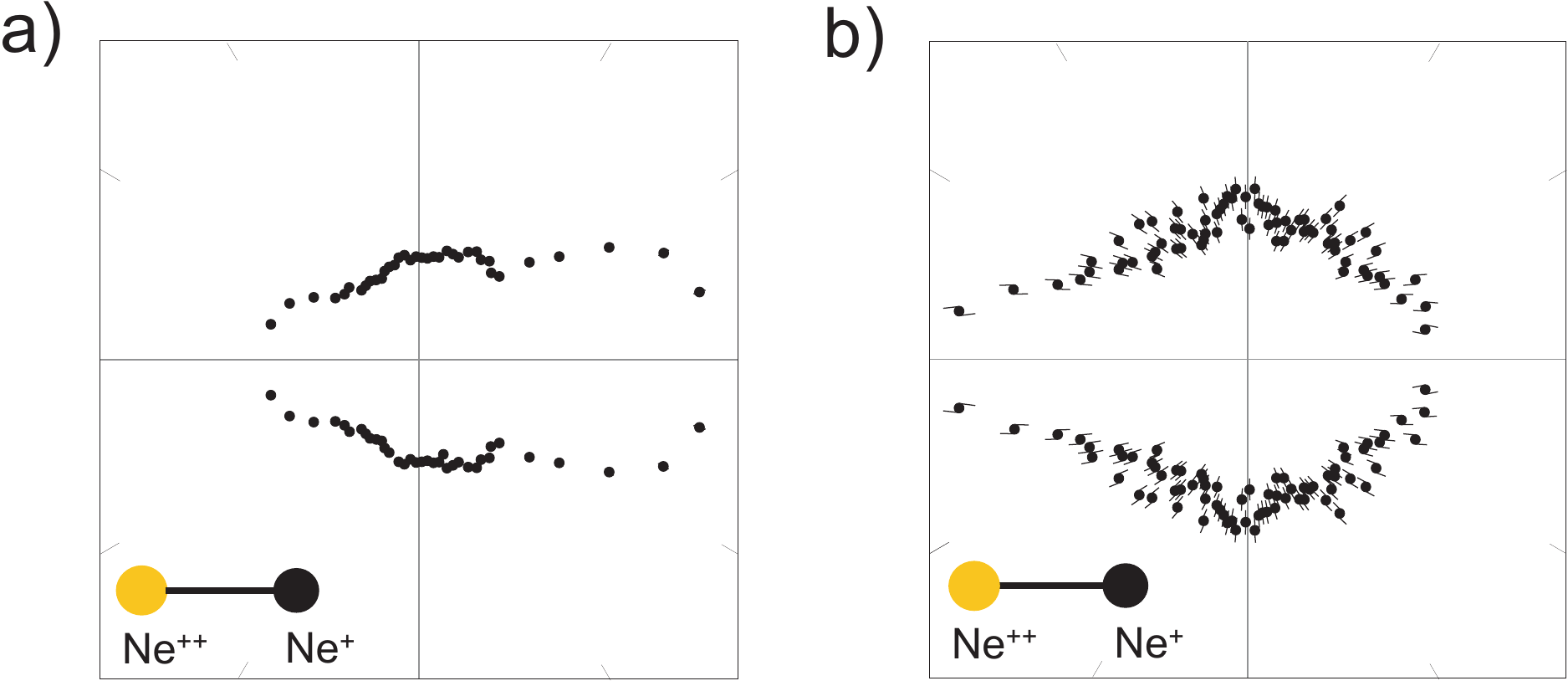}\\
  \caption{Angular distribution of the ICD-electron in the dimer frame. The dimer is aligned horizontal
  and it is integrated over the orientation of the polarization vector. a) ICD electrons which are created by
  virtual photon exchange (Channel A) in Fig.1), b) angular distribution of the ICD electrons emitted after transfer of an
  electron (Channel B in Fig.1). As suggested by the sketch in Fig.1 the direction of the
  asymmetry switches from a) to b) depending on the decay path.}
  \end{center}
  \label{Figure4}
\end{figure}

As a next step we investigate the corresponding angular distribution
for the ICD-electron. Fig. 4(a) shows the distribution for decay
path A, Fig. 4(b) the one for decay channel B. Both angular
distributions are - just as the one for the photoelectron -
asymmetric with respect to the two atomic centers of the dimer. A
striking difference is visible for the preferred direction of
emission. While the ICD electron is emitted towards the Ne$^{1+}$
for channel A) it is emitted towards the Ne$^{2+}$ in case of
pathway B). This implies the ICD electron being emitted from
opposite sites in the two cases, which is exactly what one would
intuitively expect from comparing the two processes in Fig. 1. This
finding is maybe even more surprising, as the emitted ICD-electron
originates from a valence shell: for an excited/ionized
Van-der-Waals molecule valence electrons are often viewed as
delocalized molecular orbitals.

In conclusion, the molecular frame angular distribution of
photoelectrons and electrons from interatomic Coulombic decay are
found to be strongly asymmetric for asymmetric breakup channels of
Ne$_2$. For the photoelectron this finding directly proves that a
photon induced core hole in Ne$_2$ is best thought of as being
localized. The observed asymmetry for the ICD electrons shows that
in addition to the core hole also the 2s hole created by the Auger
decay and the valence orbital that emits the ICD electron show
strong features of localization.

\acknowledgments This work was supported by BMBF, DFG, BESSY, JSPS,
and MEXT. We would like to express our thanks to Stefan Schramm and
the staff at BESSY for excellent support during the beamtime. We are
grateful also to S.~Stoychev,  A.~Kuleff, B.~Averbukh, and
L.~S.~Cederbaum for stimulating discussion. K.~K. and T.~W.
acknowledge support by DESY. X.~J.~L. acknowledges support by JSPS.

\bibliographystyle{unsrt}
%\bibliography{recoil,phot_ion,review,interference}

%\end{thebibliography}
\end{document}